# Polarization-Shaped Strong Field Control over Valley Polarization with Mid-IR Light

Igor Tyulnev[1], Julita Poborska[1], Álvaro Jiménez-Galán[2], Lenard Vamos[1], Olga Smirnova[2,3], Mikhail Ivanov[2,4,5] and Jens Biegert[1,6]

[1] ICFO - Institut de Ciencies Fotoniques, The Barcelona Institute of Science and Technology, 08860 Castelldefels (Barcelona), Spain

[2] Max-Born-Institut, Max-Born-Str. 2A, 12489 Berlin, Germany

[3] Technische Universität Berlin, Hardenbergstr. 33A, 10623 Berlin, Germany

[4] Institute für Physik, Humboldt-Universität zu Berlin, Newtonstr. 15, 12489 Berlin, Germany

[5] Department of Physics, Imperial College London, SW7 2AZ London, United Kingdom

[6] ICREA, Pg. Lluís Companys 23, 08010 Barcelona, Spain

**Abstract** We induce control over valley polarization with a polarization-shaped mid-infrared light field. The polarization state of the trefoil-shaped pump fields is measured by high harmonic spectroscopy and to confirm strong field control over non-resonant valley polarization in $MoS_2$.

## 1 Introduction

Polarization shaping of the temporal evolution of an optical light field provides additional degrees of freedom to interrogate matter and control its properties. The emergent complex electrical currents are controlled on the sub-cycle scale of the optical field and provide enticing new ways to process information or to switch the properties of a material on ultrafast timescales. The non-linear motion of field-driven electrons in a material radiate at optical frequencies; thus, detecting high-order harmonic radiation provides insight into the dynamic evolution of carriers. This potential was first recognized in Ref. [1] with the detection of high harmonics (HH) of the mid-IR driver in bulk ZnO and led to numerous investigations of the electron-hole dynamics [2,3], of many-body effects in correlated systems [4], the interplay of different bands [5], and recently the detection of quantum phases in superconductors [6]. Confining the dynamics to two-dimensional layers, e.g., in graphene-like materials or transition-metal-dichalcogenides (TMDCs), allows for the study of HH dynamics below and above the bandgap without the complication of the interpretation due to propagation effects [7]. The broken inversion symmetry of a monolayer led to observing the impact of the Berry curvature [8]. Rotation of the light's polarization vector determines the recombination times of electron-hole trajectories from K and K' symmetry points [7]. In contrast, circular polarization carries spin angular momentum that lifts the valley degeneracy through optical selection rules [9,10].

Here, we extend the control over the valley degree of freedom into the strong-field regime by matching the driving field polarization to the crystal symmetry of 2H-MoS2. To this end, we shape the optical electric field waveform's polarization in the two dimensions perpendicular to the propagation vector in a trefoil fashion. To achieve such three-fold symmetry, we combine pulses of two colors while manipulating the handedness of their circular polarization. The so-tailored field is characterized by symmetry-resolved chiral spectroscopy, where selection rules result in circularly polarized harmonics of odd and even orders from the inversion symmetric material. With the off-resonant strong-field trefoil acting as a pump, a modulation in the bandgap at high symmetry points K and K' is observed in a non-collinear geometry pump-probe experiment.

The measurement of the second harmonic of the probe field is a clear sign of the valley polarization induced by the trefoil pump field.

## 2 Valley polarization in MoS$_2$

The light carries spin angular momentum and can, thus, induce asymmetry of carriers between valleys in reciprocal space during resonant excitation of the bandgap in the K/K' points. The degree of circular polarization is linked to the valley contrasting magnetic moment "m," which defines the coupling strength $|P|^2$. For the quantum number "m" to be non-zero, the spatial inversion symmetry of the crystal must be broken. Such inversion symmetry breaking occurs naturally for monolayer TMDCs. The material becomes valley polarized, and the electron motion depends on the sign of the valley index and the Berry curvature at the respective high symmetry point [11].

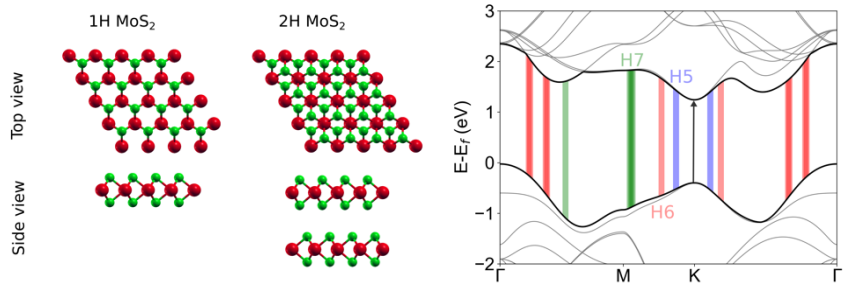

**Fig. 1.** Left: Top and side view of the trigonal prismatic crystal structure of 1H- and 2H-MoS$_2$. Right: 2H-MoS$_2$ band structure; Blue, red, and green represent gaps that fit H5, H6, and H7 emission energies and the bandwidth of the 3.2 μm driving field.

In contrast, a bi-layer system in the 2H phase is symmetric under spatial inversion due to the rotated second layer (Fig. 1), and valley asymmetry cannot be induced with circularly polarized light. However, symmetry breaking can be re-introduced through the structuring of light when the system is excited off-resonantly by a trefoil-shaped strong field. In the following, we leverage this prospect for inducing valley polarization.

## 3 Polarization shaping and trefoil generation

We use our home-built mid-infrared OPCPA system [12] at 3.2 μm for the experiment. For the polarization shaping, a 97-fs pulse is split within a Mach-Zehnder interferometer into two components in which one is frequency doubled. The polarization states and intensity ratio between fundamental and second harmonic are carefully controlled by waveplates. Both beams are recombined and then focused onto the sample. We use a vacuum electric field amplitude of 0.07 V/Å in the mid-IR and derive the probe beam at 800 nm from the same OPCPA with full optical synchronization. The pump-probe experiment is performed in non-collinear geometry to ensure the distinguishability of components from the trefoil pump and 800-nm probe.

## 4 Polarization resolved chiral spectroscopy

First, we perform chiral spectroscopy to confirm the trefoil shape of the pump beam. Analog to Ref [13], we generate HHs in GaSe with varying polarization shapes of the driving field.

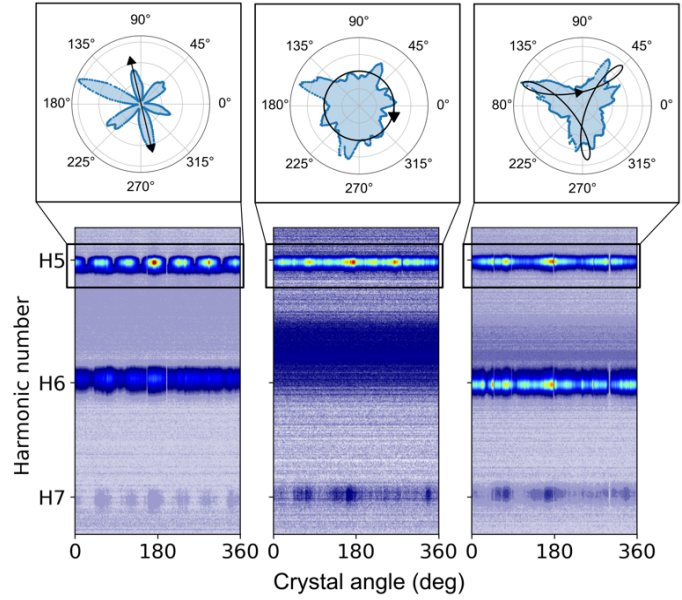

**Fig. 2.** Polarization scans for linear, circular, and trefoil driving fields. Harmonic spectra from GaSe are recorded as a function of the crystal rotation angle.

Figure 2 shows the results for the measurement of harmonic orders 5 to 7 (H5-H7). The figure shows how the harmonic signals vary with crystal rotation. Lineouts of H5 are also shown on a polar plot with the polarization type illustrated. The left column is the HH spectrum for a linearly polarized 3.2 μm driving field. The harmonic response from the hexagonal crystal yields six peaks during a full rotation scan. The harmonic intensity is maximized when the laser polarization aligns with the crystal axis (Γ-K), which occurs every 60° and completely disappears with the polarization along the Γ-M direction. Changing the driving field to circular polarization minimizes the H6 signal due to optical selection rules [14], while the other harmonics do not show any clear angle dependence due to the isotropic polarization. Finally, the two-color trefoil field shows three-fold symmetry in the H5 and H7 responses, which show maxima 120° apart. This behavior is understood when considering the crystal as a combination of Ga and Se sublattices, which are addressed by the trefoil three times each in a full rotation. Having confirmed the generation of a trefoil field, we now turn to the TMDC.

To investigate strong field-induced valley polarization, we used a 50-μm thin crystal of 2H-$MoS_2$. The material is mounted without substrate, thus avoiding contamination and background.

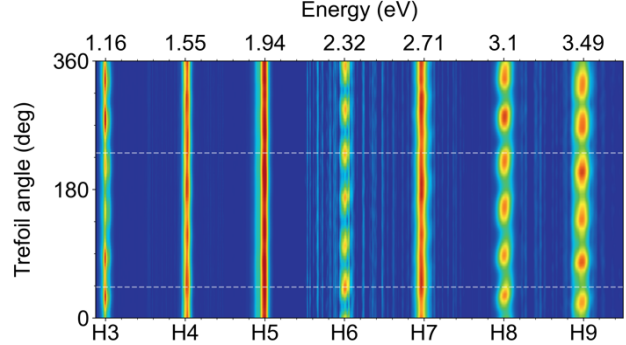

**Fig. 3.** The trefoil field drives high harmonic spectrum in 2H-MoS$_2$ as a function of the polarization rotation.

Figure 3 shows the recorded spectra for 360° of polarization rotation of the trefoil field only. Due to the six-fold symmetry of the material's 2H phase, we find that each harmonic modulates with 60° periodicity. This is in accord with scans in which the linear polarization aligns with the Γ-K direction. Note that 2H-MoS2's lowest band gap of 1.66 eV is at the K point, while the gaps at Γ and M are at 2.43 eV and 2.76 eV, respectively. From the transitions illustrated in Fig. 1, H5 is the first interband harmonic since it gets emitted closest to the K point.

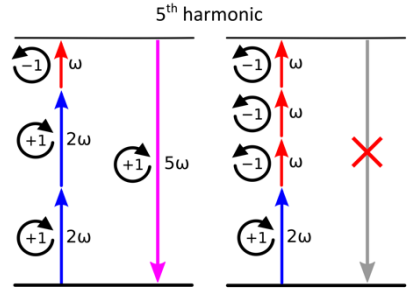

**Fig. 4.** Trefoil selection rules from two-color photon mixing with opposite spin-angular momentum.

As previously explained, although the sample is inversion symmetric, the bi-circular pump generates even harmonics. We also find that the strengths of H3 and H6 reduce by several orders of magnitude. This is explained by the selection rules (3N+1) and (3N+2) due to the conservation of spin-angular momentum during the frequency mixing of the two colors (see Fig. 4). Note that according to theory, harmonic orders (3N) are forbidden. We attribute the observable small signal to slight imperfections of the pump, i.e., slight ellipticity of the pump polarization [15]. However, the strong and clear modulation of the signal with similar peak amplitudes over the 360° rotation proves the existence of a symmetric trefoil shape.

## 5 Strong field-induced valley control

Fig. 5 shows the modulation of the second harmonic (SH) signal generated by the probe beam. A Fast Fourier Transform (FFT) of two different measurements, one with the trefoil as a pump and one without a trefoil, clearly shows that the trefoil field indeed induces non-resonant valley polarization.

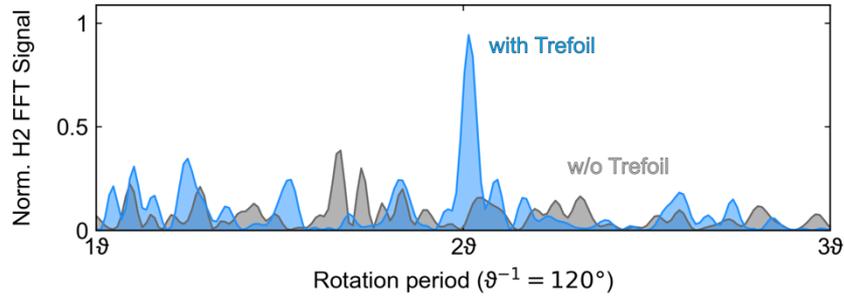

**Fig. 5.** Shown is the probe (H2) signal's dependency on the pump trefoil rotation. A clear modulation with 60° periodicity appears during strong field excitation, matching the six-fold symmetry of the 2H-MoS$_2$ sample.

When applying the trefoil field and rotating its polarization angle, the probe exhibits modulation every 60° of the pump's rotation; this shows in the FFT as frequency $2\vartheta$. As discussed in [16], strong-field interaction between the trefoil field and 2D materials like 1H-MoS2 reduces the band gap at the K or K' points, depending on the orientation of the trefoil. Note that this process is universal due to the off-resonant excitation and, thus, universal.

## 6 Summary

We show experimental results from polarization-shaped trefoil fields that non-resonantly induce valley polarization in a TMDC material. Generating and measuring the effects of valley polarization are all-optical, thus providing control over material properties at the sub-cycle scale of PHz optical fields. High harmonic spectroscopy is employed to characterize and measure the efficacy of valley polarization.

## Acknowledgments

J.B. and group acknowledge financial support from the European Research Council for ERC Advanced Grant "TRANSFORMER" (788218), ERC Proof of Concept

Grant "miniX" (840010), FET-OPEN "PETACom" (829153), FET-OPEN "OPTO-logic" (899794), FET-OPEN "TwistedNano" (101046424), Laserlab-Europe (871124), Marie Skłodowska-Curie ITN "smart-X" (860553), MINECO for Plan Nacional PID2020–112664 GB-I00; AGAUR for 2017 SGR 1639, MINECO for "Severo Ochoa" (CEX2019-000910-S), Fundació Cellex Barcelona, the CERCA Programme/Generalitat de Catalunya, and the Alexander von Humboldt Foundation for the Friedrich Wilhelm Bessel Prize.